\documentclass[12pt,a4paper]{iopart}
\usepackage{iopams} 
\usepackage{graphicx}

\usepackage[update,prepend,prefersuffix=false,suffix=]{epstopdf}

\newcommand{\bra}[1]{\langle #1 |}
\newcommand{\ket}[1]{| #1 \rangle}
\newcommand{\product}[2]{\langle #1 | #2 \rangle}

\begin{document}

\title[Maximally localized Wannier functions for 1D double-well periodic potentials]{Maximally localized Wannier functions for ultracold atoms in one-dimensional double-well periodic potentials}

\author{Michele Modugno$^{1,2}$ and Giulio Pettini $^{3}$}
\address{$^{1}$Department of Theoretical Physics and History of Science, University of the Basque Country UPV/EHU, 48080 Bilbao, Spain}
\address{$^{2}$IKERBASQUE, Basque Foundation for Science, 48011 Bilbao, Spain}
\address{$^{3}$Dipartimento di Fisica e Astronomia, Universit\`a di Firenze,
and INFN, 50019 Sesto Fiorentino, Italy}
\ead{michele\_modugno@ehu.es, pettini@fi.infn.it}

\begin{abstract}
We discuss a method for constructing generalized Wannier functions that are maximally localized at the minima of a one-dimensional periodic potential with a double-well per unit cell. By following the approach of (Marzari M and Vanderbilt D 1997 \textit{Phys. Rev. B} \textbf{56}, 12847), we consider a set of band-mixing Wannier functions with minimal spread, and design a specific two-step gauge transformation of the Bloch functions for a composite two band system. This method is suited to efficiently computing the tight-binding coefficients needed for mapping the continuous system to a discrete lattice model. Their behaviour is analyzed here as a function of the symmetry properties of the double-well (including the possibility of parity-breaking), in a range of feasible experimental parameters.
\end{abstract}

\date{\today}



\section{Introduction}
\label{sec:intro}

Ultracold atoms in optical lattices are attracting an increasing interest as quantum simulator for condensed matter systems \cite{bloch,lewenstein}. 
Optical lattices are realized by means of one or more continuous sinusoidal potentials created by exploiting the electric dipole interactions between the atoms and the laser beams \cite{bloch}.
Depending on the beam geometry one can realize one-, two-, or three-dimensional periodic lattices, with one or more wells per unit cell \cite{yukalov}; quasiperiodic structures are also possible (see e.g. \cite{lewenstein,santos,fallani,roati,modugno}).

In the tight binding regime, when the lattice intensity is sufficiently high so that the atoms are deeply localized in the lowest vibrational states of the potential wells (each well being associated to a site of a discrete lattice), it is convenient to map the system hamiltonian onto a discrete lattice model (or \textit{tight binding model}), representing the usual tool for theoretical calculations. 
The paradigms are the Hubbard model for fermions \cite{hubbard}, and the Bose-Hubbard model for bosons \cite{fisher}.
These models are characterized by tunnelling coefficients related to the hopping between neighbouring sites, and interaction strengths which characterize the onsite interaction among the atoms, whose actual values depend on the parameters of the underlying continuous model. 

As the mapping between the continuous and discrete versions of the system hamiltonian is achieved by means of an expansion over a basis of localized functions at each potential well \cite{bloch,jaksch,zwerger}, a precise knowledge of these basis functions is therefore important to connect the actual experimental parameters with the coefficients of the discrete model employed in the theoretical calculations.

For a simple sinusoidal potential, the natural basis is provided by the exponentially decaying Wannier functions discussed by Kohn \cite{wannier,kohn}. Notably, in this case the expression for the tunnelling coefficient turns out to depend just on the Bloch spectrum \cite{he}, being therefore independent on the basis choice (this does not hold for the interaction coupling). Analytic expressions for both coefficients can be obtained by means of different approximations \cite{zwerger,gerbier}.

In the case of two wells per unit cell, the Kohn-Wannier recipe is not sufficient. For example, for a symmetric double-well the Kohn-Wannier functions display the same symmetry as the local potential structure \cite{kohn,cloiz} and thus they occupy both wells and cannot be associated to a single lattice site.
A common approach used in the literature is that of the so-called atomic orbitals \cite{ashcroft,wallace,reich}, that has been recently employed e.g. for the case of a symmetric double-well unit cell of two-dimensional graphene-like optical lattices \cite{miniatura}. This method is based on a specific ansatz, according to which tight-binding Wannier functions are constructed from linear combinations of wave functions deeply localized in the two potential wells of the unit cell.

A more general approach is the one proposed by Marzari and Vanderbilt \cite{marzari},
where maximally localized Wannier functions (MLWFs) are obtained by minimizing the spread of a set of  Wannier functions by means of a suitable gauge transformation of the Bloch eigenfunctions. This method coincides with the Kohn method for the single band MLWFs of a one-dimensional potential, but it can be extended to more complex situations when \textit{generalized} MLWFs for composite bands are needed. The method is implemented by means of a software package, and is largely employed for computing MLWFs of real condensed matter systems \cite{wannier90}.

In this article, we consider the case of a one-dimensional periodic potential with a double-well per unit cell \cite{trotzky,sebby,trebst,danshita,barmettler,qian1,qian2}, discussing an alternative method for constructing the low-lying generalized MLWFs based on the minimal spread requirement of Marzari \textit{et al.} \cite{marzari}, specifically suited for double-well potentials.
In particular, we consider a composite band formed by the two lowest Bloch bands and, differently from \cite{marzari}, we design a two-step gauge transformation specific for a composite two band system, that can be solved by integrating a set of ordinary differential equations, with suitable boundary conditions. This allows to efficiently compute the tunnelling coefficients and the other tight binding coefficients in terms of the parameter of the continuous potential. We also remark that, though the approach of Marzari \textit{et al.} \cite{marzari} was proposed for degenerate bands, we find that the method works properly in a wider regime, provided that the first band gap does not exceed the distance between the second and third band.

The paper is organized as follows. In Section \ref{sec:tb} we review the mapping of a many-body hamiltonian onto a tight binding model, discussing in particular the case of a sinusoidal periodic potential with a double well in the unit cell. Then, in Section \ref{sec:mlwfs} we describe the gauge transformation for constructing the \textit{generalized} MLWFs, giving some examples and comparing them with those of the single band approach. Then, in Section \ref{sec:regimes}, we discuss the regime of validity of the \textit{composite band} approach, and compare the predictions  of the \textit{full} and \textit{nearest-neighbour} versions of the model  by discussing the behaviour of the tunnelling coefficients and the other tight-binding parameters. Final considerations are drawn in the Conclusions. Technical points regarding the mapping on momentum space and the numerical implementation are discussed in the Appendices.

\section{Tight binding models for double well optical lattices}
\label{sec:tb}
Let us start by reviewing how tight-binding models are defined from a continuous potential.
As a specific example, here we consider a one-dimensional many-body hamiltonian for 
ultracold bosons \cite{jaksch}
\begin{equation} 
\hat{\cal{H}} = \int\!d x \;\hat{\psi}^\dagger\hat{H}_{0}\hat{\psi} + 
\frac{g}{2}\int\! d x \,\hat{\psi}^\dagger\hat{\psi}^\dagger\hat{\psi}\hat{\psi}\equiv\hat{\cal{H}}_0+\hat{\cal{H}}_{int}
\label{eq:mbham}
\end{equation}
with $\hat\psi(x)$ being the bosonic field operator, $\hat{H}_{0}=-({\hbar^2}/{2m})\nabla^2+V (x)$ the single particle hamiltonian, and $V(x)$ a double-well periodic potential of the form 
\begin{equation}
V(x)=V_{1}\sin^{2}\left(k_{B}x +\phi_{0}\right)+ V_{2}\sin^{2}\left(2k_{B} x+\theta_{0}+2\phi_{0}\right)
\label{eq:potential}
\end{equation}
with $k_{B}=\pi/d$ and $V_{1}$ strictly non vanishing ($V_{1}>0$) in order to fix the overall period to $d$, $V(x+d)=V(x)$. This is a typical potential used in the experiments with ultracold atoms \cite{bloch}, by which one can construct an \textit{optical lattice} with one or two wells in the unit cell. In particular, the condition for having two minima in each period for any value of $\theta_{0}$ is $V_{2}>0.5V_{1}$. 
In the following, the potential amplitudes $V_{i}$ will be expressed in units of $E_{R}={\hbar}^2k_{B}^2/2 m$, the so-called \textit{recoil energy} for an atom 
absorbing a photon of the first lattice.
\begin{figure}
\centerline{\includegraphics[width=0.75\columnwidth]{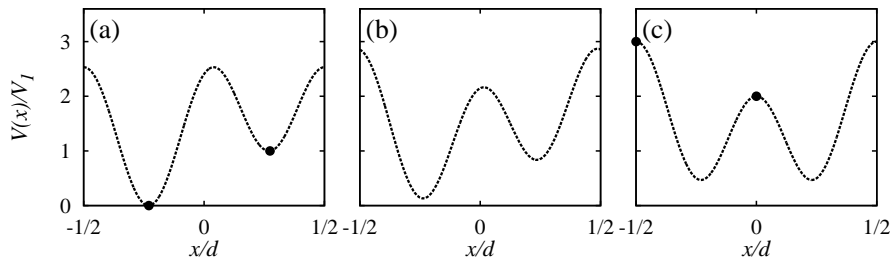}}
\caption{The three possible configurations for the unit cell of the potential in  (\ref{eq:potential}) (here $V_{2}=2V_{1}$): (a) two different minima, with the overall potential having two centers of parity, for $\theta_{0}=0$ ($\phi_{0}\simeq \pi/4$); (b) an asymmetric double-well with parity that is broken globally - in this example $\theta_{0}=\pi/4$ ($\phi_{0}\simeq\pi/8$); (c) a symmetric double-well, $\theta_{0}=\pi/2$ ($\phi_{0}=0$). The black dots in (a), (c) represent the parity centers of the whole periodic potential.}
\label{fig:pot}
\end{figure}

The phases $\theta_{0}$ and $\phi_{0}$ are arbitrary; $\phi_{0}$ represents a rigid shift of the whole potential, while $\theta_{0}$ can be varied along with the ratio of the amplitudes $V_{2}/V_{1}$ to change the landscape of the potential.
For convenience, the unit cell is defined as having two maxima at the cell borders, with both minima inside the cell (see figure \ref{fig:pot}). In addition, the angle $\phi_{0}$ is tuned in order to have a unit cell centered in $x=0$, $x\in[-d/2,d/2]$, with the absolute minimum in the left well \footnote{The presence of $\phi_{0}$ is also useful for testing the robustness of the numerical method for obtaining the MLWFs, that indeed should not depend on an overall translation of the potential.}. 
Depending on the value of $\theta_{0}$, it is possible to realize three different configurations, as shown in figure \ref{fig:pot}. 
(a) A unit cell with two different minima and with degenerate maxima, for $\theta_{0}=n\pi$ ($n\in\mathbb{Z}$); in this case the whole periodic potential has two (classes of) parity centers, corresponding to the two minima, with all the maxima being degenerate. 
(b) An asymmetric double-well with parity that is globally broken, for any $\theta_{0}\in(0,\pi/2)+n\pi/2$.
(c) A symmetric double-well in the unit cell, for $\theta_{0}= \pi/2+n\pi$; in this case the potential has again two centers of parity, now at the two maxima, with all the minima being degenerate.

As anticipated in the Introduction, when the potential wells are deep enough, it may be convenient to map the hamiltonian (\ref{eq:mbham}) onto a \textit{tight binding model} on the discrete lattice corresponding to the potential minima, by expanding the field operator $\hat{\psi} (x)$ on a basis $\{f_{nj}(x)\}$ of functions localized around each minimum
\begin{equation} 
\label{eq:sviluppo}
\hat\psi (x)\equiv\sum_{nj}\hat{a}_{nj}f_{nj}(x)
\end{equation}
where $\hat{a}_{nj}^{\dagger}$ ($\hat{a}_{nj}$) represent the creation (destruction) operator of a single particle at site $j$, and satisfy the usual commutation rules $[\hat{a}_{nj},\hat{a}^{\dagger}_{n'j'}]=\delta_{jj'}\delta_{nn'}$ (following from those for the field $\hat{\psi}$). 

In the presence of a single well per unit cell, it is known that a basis of localized functions is provided by the exponentially decaying Wannier functions $w_{nj}(x)$ discussed by Kohn \cite{kohn,he}. 
In general, this is not the case when there are two wells per unit cell. For example, for a symmetric double-well the Kohn-Wannier functions display the same symmetry of the local potential structure \cite{kohn,cloiz} and thus they cannot be associated to a single lattice site as as they occupy both wells in the unit cell.
In the next section we will show that when the two lowest Bloch bands are sufficiently close to each other with respect
to the third band (as it will be clear from the discussion in Sect. \ref{sec:examples} and \ref{sec:regimes}), we can construct a set of \textit{generalized} Wannier functions $\tilde{w}_{nj}(x)$ that are maximally localized at each minima, by following the approach of Marzari and Vanderbilt \cite{marzari} for a \textit{composite band}. This corresponds to the generalization of the \textit{single band approximation} (in case of a single well lattice) to the double well case, as we need at least two localized functions in each lattice cell to map the system on the discrete lattice. Then, in Section \ref{sec:regimes}, we will explore the range of validity of this composite band approaches, highlighting the different implications on the structure of different tight binding models.

\begin{figure}
\centerline{\includegraphics[width=0.6\columnwidth]{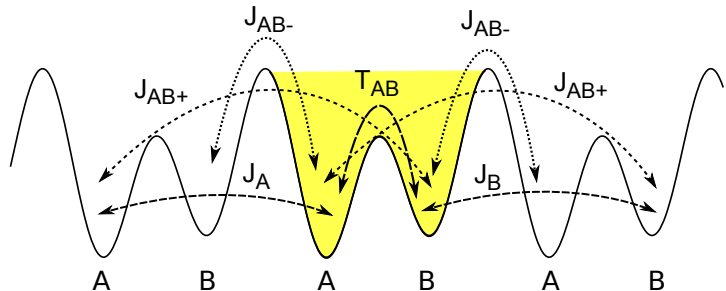}}
\caption{A sketch of the double-well structure and of the tunnelling coefficients between sites A and B.}
\label{fig:doublewell}
\end{figure}

In the following we will restrict the analysis to the two lowest energy bands, in analogy with the single band approximation for the Bose-Hubbard model \cite{zwerger}.
Then, within this approximation, the single particle hamiltonian can be written as
\begin{equation}
\hat{\cal{H}}_{0} \simeq \sum_{\nu\nu'=A,B}\sum_{jj'}\hat{a}_{j\nu}^\dagger\hat{a}_{j'\nu'}
\langle {f}_{j\nu}|\hat{H}_{0}|{f}_{j'\nu'}\rangle
\label{eq:full-Hcb}
\end{equation} 
where $j$ is the unit cell index whereas $\nu=A,B$ substitutes the band index $n=1,2$ being an internal index labelling the left and right sub-wells respectively (see figure \ref{fig:doublewell}).
Here the expansion coefficients correspond to the onsite energies $E_{\nu}=\langle {f}_{j_{\nu}}|\hat{H}_{0}|{f}_{j_{\nu}}\rangle$, and to the tunnelling amplitudes 
between different (sub)wells $T_{\nu\nu'}^{jj'}\equiv-\langle {f}_{j\nu}|\hat{H}_{0}|{f}_{j'\nu'}\rangle$. 
In general, it is customary to further approximate the above expression by neglecting the coupling beyond nearest neighbours both for the single-well \cite{jaksch} and double-well lattices \cite{trebst,qian1,qian2}. This is a reasonable assumption for a single well lattice in the tight binding regime \cite{he,boers}, but may not be fully justified
in the range of the typical experimental parameters for a double well, as we will discuss later on. For this reason, here we keep all the terms corresponding to nearest-neighbouring cells, characterized by the following tunnelling coefficients 
\begin{eqnarray}
\label{eq:jnu}
J_{\nu}&=&-\langle {f}_{j_{\nu}}|\hat{H}_{0}|{f}_{(j+1)_{\nu}}\rangle
\\
T_{AB}&=&-\langle {f}_{j_{A}}|\hat{H}_{0}|{f}_{j_{B}}\rangle
\\
J_{AB_{\pm}}&=&-\langle {f}_{j_{A}}|\hat{H}_{0}|{f}_{(j\pm1)_{B}}\rangle
\label{eq:jab}
\end{eqnarray}
as shown in figure \ref{fig:doublewell}.
The corresponding hamiltonian is 
\begin{eqnarray}
\hat{\cal{H}}_{0} &\simeq 
\sum_{\nu=A,B}\sum_{j}E_{\nu}\hat{n}_{j_{\nu}}
-\sum_{\nu=A,B}\sum_{j}J_{\nu}(\hat{a}_{j_{\nu}}^\dagger\hat{a}_{(j+1)_{\nu}} +h.c.)
\nonumber\\
&
-\sum_{j}\left(T_{AB}\hat{a}_{j_{A}}^\dagger\hat{a}_{j_{B}}+
J_{AB_{+}}\hat{a}_{j_{A}}^\dagger\hat{a}_{(j+1)_{B}} +
J_{AB_{-}}\hat{a}_{j_{A}}^\dagger\hat{a}_{(j-1)_{B}} +h.c.\right).
\label{eq:fullH}
\end{eqnarray} 

When the tunnelling $J_{\nu}$ and $J_{AB_{+}}$ are negligible the above
expression can be further simplified by retaining just the coupling between nearest neighbouring wells 
\begin{eqnarray}
\hat{\cal{H}}_{0} &\simeq 
\sum_{\nu=A,B}\sum_{j}E_{\nu}\hat{n}_{j_{\nu}}
-
\sum_{j}\left(T_{AB}\hat{a}_{j_{A}}^\dagger\hat{a}_{j_{B}}+
J_{AB_{-}}\hat{a}_{j_{A}}^\dagger\hat{a}_{(j-1)_{B}}
+h.c.\right).
\label{eq:H0-cb}
\end{eqnarray} 
This is the analogous of the nearest-neighbours approximation for the single-well case and it is commonly used in the literature \cite{qian1,qian2}. Hereinafter, we will refer to the approximate hamiltonians in (\ref{eq:fullH}) and (\ref{eq:H0-cb}) as the (single particle) \textit{full} and \textit{nearest-neighbour} tight-binding models, respectively.

Finally, the general form of the interaction term is
\begin{equation}
\hat{\cal H}_{int} = \frac{g}{2}\sum_{\{\nu_{i}\}=A,B}
\sum_{\{j_{i}\}}\hat{a}_{j_{1}\nu_{1}}^\dagger\hat{a}_{j_{2}\nu_{2}}^\dagger\hat{a}_{j_{3}\nu_{3}}\hat{a}_{j_{4}\nu_{4}}
\int_{x}f_{j_{1}\nu_{1}}^\ast f_{j_{2}\nu_{2}}^\ast
f_{j_{3}\nu_{3}}f_{j_{4}\nu_{4}}.
\label{eq:Hint-full}
\end{equation}
Here we consider just the onsite contributions in the two sub-wells A and B
\begin{equation}
\hat{\cal H}_{int}\simeq \sum_{\nu=A,B}\frac{U_{\nu}}{2}\sum_{j_{\nu}}\hat{n}_{j_{\nu}}\left(\hat{n}_{j_{\nu}}-1\right)
\label{eq:Hint-cb}
\end{equation}
with $U_{\nu}=g\int\!d x \left|{f}_{j_{\nu}} (x)\right|^4$. This corresponds to the usual approximation used for the single well case (see e.g. \cite{jaksch}), and is justified when the basis functions $\{f_{nj}(x)\}$ are strongly localized in each sub-well. In general, once these functions are known, the next-to-leading interaction terms can be straightforwardly computed by evaluating the corresponding superposition integral of four $f_{nj}$'s.

\subsection{Single particle spectrum}
\label{sec:spectrum}

Let us now consider the single particle spectrum of the hamiltonian (\ref{eq:fullH}). 
By defining
\begin{equation}
\hat{b}_{\nu{k}}=\sqrt{\frac{d}{2\pi}}
\sum_{j} ~e^{i{k}{ja}}\hat{a}_{j{\nu}}
\end{equation}
the \textit{full} tight binding hamiltonian (\ref{eq:fullH}) can be rewritten as
\begin{equation}
\hat{\cal{H}}_{0}=\sum_{\nu\nu'}\int_{\cal B} d{k}~h_{\nu\nu'}(k)\hat{b}_{k\nu}^{\dagger}\hat{b}_{k\nu'}
\end{equation}
with
\begin{equation}
h_{\nu\nu'}(k)=\left(\begin{array}{cc}
 \epsilon_{A}(k) & Z(k) \\
 Z^{*}(k) & \epsilon_{B}(k)
\end{array}\right)
\end{equation}
and
\begin{eqnarray}
\epsilon_{\nu}(k)&\equiv E_{\nu}-2J_{\nu}\cos(kd)
\\
Z(k)&\equiv -(T_{AB} + J_{AB_{+}}e^{-ikd} +J_{AB_{-}}e^{ikd}),
\end{eqnarray}
where the operators $\hat{b}_{k\nu}$ satisfy the canonical commutation relations, $[\hat{b}_{\nu{k}},\hat{b}_{\nu'{k}'}^{\dagger}]=\delta({k}-{k'})\delta_{\nu\nu'}$.
Then, by diagonalizing the matrix $h(k)$, and defining $\epsilon_{\pm}(k)\equiv(\epsilon_{A}(k)\pm \epsilon_{B}(k))/2$, we get 
(see also \cite{miniatura})
\begin{equation}
\varepsilon_{\pm}^{tb}(k)=\epsilon_{+}(k)\pm\sqrt{\epsilon_{-}^{2}(k)+|Z(k)|^{2}}
\label{eq:ener-tb}
\end{equation}
that represents the spectrum of the \textit{full} tight binding model in  (\ref{eq:fullH}). In addition, with $J_{\nu}=0=J_{AB+}$, the same expression gives also the spectrum for the \textit{nearest-neighbour} approximation in  (\ref{eq:H0-cb}).

Instead, in the \textit{single band} case we simply have \cite{bloch}
\begin{equation}
\label{eq:ener-sb}
\varepsilon_{n}^{sb}(k)= E_{n}^{sb}-2J_{n}^{sb}\cos(kd)
\end{equation}
with
\begin{equation}
E_{n}^{sb}={\frac{a}{2\pi}}\int_{\cal B}\!\!d{k} ~\varepsilon_{n}(k);\qquad
J_{n}^{sb}=-{\frac{a}{2\pi}}\int_{\cal B}\!\!d{k} ~\varepsilon_{n}(k) e^{ika}
\end{equation}
$\varepsilon_{n}(k)$ being the exact Bloch spectrum; notably these expressions do not depend on the choice of the Wannier basis.

\section{Generalized Wannier functions}
\label{sec:mlwfs}

In this section we discuss the method for constructing the MLWFs for the double well case. 
In order to fix the notations, let us first recall some basic properties of periodic systems \cite{ashcroft,callaway}. 
Owing to the Bloch's theorem, the eigenfunctions of the single particle hamiltonian $\hat{H}_{0}$ can be written as $\psi_{nk}(x)=e^{ikx}u_{nk}(x)$, the $u_{nk}(x)$'s having the same periodicity of the potential and satisfying 
the following normalization in the unit cell, $\product{u_{mk}}{u_{nk}}=({d}/{2\pi})\delta_{mn}$. 
The Wannier functions for a single band are defined as
\begin{equation}
w_n({x}-{R}_j)=\sqrt{\frac{d}{2\pi}}
\int_{\cal B} d{k} ~e^{-i{k}{R}_{j}}\psi_{nk}(x)\equiv w_{nj}(x)
\end{equation} 
with ${\cal B}$ indicating the first Brillouin zone, $k\in[-k_{B},k_{B}]$, and $R_{j}\equiv jd$, whereas generalized Wannier
functions for composite bands have the same formal expression but are built up from a linear combination of Bloch eigenstates, namely
\begin{eqnarray}
\label{eq:wanniergen}
{\tilde w}_n({x}-{R}_j)&=&\sqrt{\frac{d}{2\pi}}
\int_{\cal B} d{k} ~e^{-i{k}{R}_{j}}\sum_{m}U_{nm}(k)\psi_{mk}(x)
\\
&\equiv&\sqrt{\frac{d}{2\pi}}
\int_{\cal B} d{k} ~e^{-i{k}{R}_{j}}{\tilde\psi}_{nk}(x)
\label{eq:mlwfs}
\end{eqnarray}
with $UU^{\dagger}=1$ and $U_{nm}(k+2k_{B})=U_{nm}(k)$. The Wannier functions satisfy the ortho-normality relation
$\product{w_{nj}}{w_{n'j'}}=\product{{\tilde w}_{nj}}{{\tilde w}_{n'j'}}=\delta_{nn'}\delta_{jj'}$. 
We also remark that the generalized Bloch functions ${\tilde\psi}_{nk}$ are not eigenstates of $\hat{H}_{0}$; however, their ortho-normality relations are preserved, owing to the unitarity of the transformation matrices $U_{nm}(k)$.

Different choices of the matrices $U_{nm}(k)$ lead to different results and it is customary to speak about
{\it gauge dependence} of the Wannier functions due to the $k$-dependence of the $U(n)$ transformations. In the single band case
the arbitrariness reduces to the abelian $U(1)$ group of phase transformations coming from the freedom in choosing the Bloch basis at each $k$. 

A general approach to obtain MLWFs has been proposed in a seminal paper by Marzari and Vanderbilt \cite{marzari},
 where MLWFs are obtained by minimizing the generalized Wannier spread $\Omega=\sum_{n}\left[\langle x^2\rangle_{n}-\langle x\rangle_{n}^{2}\right]$
 by means of a suitable gauge transformation of the Bloch eigenfunctions. In the case of a single band
 the method returns the Kohn result \cite{marzari}. Marzari and Vanderbilt have shown that the spread can be written
 as the sum of two positive terms, $\Omega=\Omega_{I}+\bar{\Omega}$, the first being gauge invariant and therefore fixing the minimal spread. 
The gauge dependent term $\bar{\Omega}$ can be further split into the diagonal and off-diagonal components,
 ${\bar\Omega}=\Omega_D+\Omega_{OD}$, that in the one-dimensional case read
\begin{eqnarray}
\Omega_D &=&\sum_{n}\sum_{j\ne0}\Big|\langle w_{nj}|{\hat x}|w_{n0}\rangle\Big|^2
\\
\Omega_{OD}&=&\sum_{m\neq n}\sum_j
\Big|\langle w_{mj}|{\hat x}|w_{n0}\rangle\Big|^2.
\end{eqnarray}

Both $\Omega_{D}$ and $\Omega_{OD}$ can be written in terms of the generalized Berry vector potentials $A_{nm}(k)$, defined as \cite{niu1,pettini}
\begin{equation}
A_{nm}(k)=i\frac{2\pi}{d}\bra{u_{nk}}\partial_{k}\ket{u_{mk}}
\end{equation}
with the matrix $A(k)$ being hermitian, $A^{\dagger}(k)=A(k)$ (it follows from $\partial_{k}\product{u_{nk}}{u_{mk}}=0$).
We also recall that the integral over the first Brillouin zone of $A_{nn}(k)$ gives the one band Zak-Berry phases $\gamma_{n}$ 
\begin{equation}
\label{eq:zakphase}
\gamma_{n}=i\frac{2\pi}{d}\int_{\cal{B}}\bra{u_{nk}}\partial_{k}\ket{u_{nk}}=\int_{\cal{B}}A_{nn}(k)\equiv\frac{2\pi}{d}
\langle A_{nn}\rangle_{\cal{B}}
\end{equation}
which are proportional to the offset of the Wannier function centers, 
$\langle x\rangle_{n0}=
\left(\langle x\rangle_{nj} - R_{j}\right)=(d/2\pi)\gamma_{n}$ \cite{bohm,marzari}, 
yielding $\langle x\rangle_{n0}=\langle A_{nn}\rangle_{\cal{B}}$ (this relation is preserved under a generic unitary gauge transformation, as in (\ref{eq:mlwfs})).
It is also worth to remember that the single Wannier centers are invariant only under single band $U(1)$ gauge transformations, whereas for general
$U(n)$ transformations only their sum is conserved \cite{marzari}. 

Then, the expressions for $\Omega_{D}$ and $\Omega_{OD}$ can be written as
\begin{eqnarray}
\label{eq:omegad}
\Omega_D&=&\sum_{n}\langle \left(A_{nn}(k)-\langle A_{nn}\rangle_{\cal{B}}\right)^{2}\rangle_{\cal{B}}=\sum_{n}\Omega_{Dn}
\\
\Omega_{OD}&=&\sum_{m\neq n}\langle |A_{nm}|^{2}\rangle_{\cal{B}}\,,
\label{eq:omegaod}
\end{eqnarray}
and in general can be reduced by means of a functional minimization in $k$ space, as discussed in \cite{marzari,wannier90}.

Here we use a different approach, specifically suited for constructing the set of MLWFs for the double-well case. Namely, we show that the minimization problem can be reformulated by identifying a specific gauge transformation for a composite band in one dimension, expressed in terms of a set of ordinary differential equations with periodic boundary conditions. We recall that in one dimension ${\tilde\Omega}$ can be made strictly vanishing, and this corresponds to find a gauge (also called \textit{parallel transport} gauge) in which the matrix $A_{nm}(k)$ is diagonal, with the diagonal elements being constant and equal to their mean values. The latter are related to the eigenvalues of the matrix generalizing the Berry phase to the non abelian case \cite{marzari}.

\subsection{Gauge transformations and differential equations}

The diagonal and off-diagonal spreads $\Omega_{D}$ and $\Omega_{OD}$ can be minimized either simultaneously or independently.
For the following discussion, it is useful to distinguish between two kind of gauge transformations.

I. Diagonal $U(n)$ transformations which correspond to a set of \textit{single band} gauge transformations of the form
\begin{equation}
\label{eq:gauge1}
|u_{nk}\rangle\rightarrow |\tilde{u}_{nk}\rangle=e^{ i\phi_{n}(k)} |u_{nk}\rangle
\end{equation}
with $\phi_{n}(k)$ being a real continuous (differentiable) function of $k$, such that $\phi_{n}(k+2k_{B})=\phi_{n}(k) + 2\pi \ell$  ($\ell$ integer) in order to have periodic
and single valued Bloch eigenstates. Then, we can set $\ell=0$ without loss of generality. 
As discussed in \cite{marzari}, this transformation may be used to minimize each term of $\Omega_{D}$, as it affects 
the Wannier spread $\Omega_n=\langle x^2\rangle_{n}-\langle x\rangle_{n}^{2}$ while preserving their centers $\langle x\rangle_n$ (modulo a lattice vector). 
In particular, $\Omega_{D}$ can be set exactly to zero \cite{marzari}.
In fact, we have $A_{nn}(k)\rightarrow A_{nn}(k)-{\partial_{k} \phi_{n}}(k)$, and therefore
\begin{equation}
\Omega_{Dn}\rightarrow \tilde{\Omega}_{Dn}=\langle \left(A_{nn}-{\partial_{k} \phi_{n}} -\langle A_{nn}\rangle_{\cal{B}}\right)^{2}\rangle_{\cal{B}}
\end{equation}
that vanishes by imposing
\begin{equation}
{\partial_{k} \phi_{n}}=A_{nn}-\langle A_{nn}\rangle_{\cal{B}}\,.
\label{eq:gauge-a}
\end{equation}
This equation can be readily solved numerically, as discussed in \ref{sec:numerics} and \ref{sec:gaugetransf}.
In addition, it is straightforward to verify that under the transformation (\ref{eq:gauge1}), $A_{12}(k)$ changes only by 
a phase factor, and therefore $\Omega_{OD}$ remains unchanged.

II. A \textit{full gauge transformation in the composite band} of the form
\begin{equation}
|u_{nk}\rangle\rightarrow |\tilde{u}_{nk}\rangle=\sum_{m}U_{nm}(k)|u_{mk}\rangle
\label{gaugematrix}
\end{equation}
where, as already said, $U(k)\in U(N)$ (for a $N$-composite band), constrained to the following periodic condition
\begin{equation}
U_{nm}\left(k+2k_{B}\right)=U_{nm}(k)\,.
\label{eq:u-periodicity}
\end{equation}

Under such a transformation the generalized Berry potentials transform as 
\begin{eqnarray}
A_{nm}\rightarrow{\tilde A}_{nm}&=&i\frac{2\pi}{a}\int dx~ {\tilde u}^{*}_{n}{\partial_{k} {\tilde u}_{mk}}
\nonumber\\
&=&i\sum_{l}U^{*}_{nl}{\partial_{k} U_{ml} }+\sum_{l,l'}U^{*}_{nl}U_{ml'}A_{ll'}
\label{eq:atilde}
\end{eqnarray}

In general, $U(N)$ can be written as a semidirect product $SU(N)\rtimes U(1)$, with $U(1)$ being subgroup of $U(N)$ 
consisting of matrices of the form diag$(1,1,\dots,e^{i\chi})$. 
As anticipated, here we restrict to a $2\times2$ case, and therefore we have
\begin{equation}
U(k) = \left(\begin{array}{cc}
 z_{1}(k) & -z^{*}_{3}(k) \\
 z_{3}(k) & z^{*}_{1}(k)
 \end{array} \right)
 \left(\begin{array}{cc}
 1 & 0 \\
 0 & r(k)
 \end{array} \right) 
\label{eq:u0matrix}
\end{equation}
with $|z_{1}|^{2} + |z_{3}|^{2}=1$, $r(k)=e^{{i\chi(k)}}$. Moreover, by using the following parametrization for a matrix $S\in SU(2)$, $S=e^{\displaystyle i\alpha\vec{\sigma}\cdot \hat{n}/2}$ with $\hat{n}=(\cos\varphi\sin\theta,\sin\varphi\sin\theta,\cos\theta)$ 
and $\sigma_i$ being the Pauli matrices, we can write
\begin{equation}
U =
 \left(\begin{array}{cc}
\cos\frac{\alpha}{2}+i\sin\frac{\alpha}{2}\cos\theta~ & ie^{\displaystyle{i(\chi-\varphi)}}\sin\theta\sin\frac{\alpha}{2} \\
ie^{\displaystyle{i\varphi}}\sin\theta\sin\frac{\alpha}{2} & e^{\displaystyle{i\chi}}\left(\cos\frac{\alpha}{2}-i\sin\frac{\alpha}{2}\cos\theta\right) 
 \end{array} \right)
 \label{eq:umatrix}
\end{equation}
with $\chi=\chi(k)$, $\varphi=\varphi(k)$, $\alpha=\alpha(k)$ and $\theta=\theta(k)$. 

Then, since $\Omega$ transforms as follows
\begin{equation}
\Omega_{Dn}\rightarrow \tilde{\Omega}_{Dn}=
\langle (\tilde{A}_{nn}(k)-\langle \tilde{A}_{nn}\rangle_{\cal{B}})^{2}\rangle_{\cal{B}}
\end{equation}
\begin{equation}
\Omega_{OD}\rightarrow \tilde{\Omega}_{OD}=2\langle |\tilde{A}_{12}|^{2}\rangle_{\cal{B}}
\end{equation}
in order to get $\tilde{\Omega}=0$ one has to impose (see  (\ref{eq:atilde}))
\begin{eqnarray}
\label{eq:ann}
A_{nn}(k)&\equiv& i\sum_{l}U^{*}_{nl}{\partial_{k} U_{nl} }+\sum_{l,l'}U^{*}_{nl}U_{nl'}A_{ll'}=\langle {\tilde A}_{nn}\rangle_{{\cal B}}\qquad(n=1,2)
\\
\tilde{A}_{12}(k)&\equiv& i\sum_{l}U^{*}_{1l}{\partial_{k} U_{2l} }+\sum_{l,l'}U^{*}_{1l}U_{2l'}A_{ll'}=0.
\label{mastereq}
\end{eqnarray}
Notice that the former,  (\ref{eq:ann}), is an integro-differential equation where
the right-end side corresponds to the center of the Wannier functions, $\langle {\tilde A}_{nn}\rangle_{{\cal B}}=\langle \tilde{w}_{nj}|{\hat x}|\tilde{w}_{nj}\rangle$, that are not known \textit{a priori} (only their sum is conserved in the {\it parallel transport} gauge).

Therefore, in the following we will consider a specific transformation that makes $\Omega_{OD}$ vanish without specific requirements on the transformation of $\Omega_{D}$, given by the solution of  (\ref{mastereq}).
Then, by using  (\ref{eq:umatrix}), the latter can be transformed in a system of four differential equations for $\alpha$, $\theta$, $\varphi$, $\chi$, whose normal form is
\begin{eqnarray}
\label{eq:alpha}
\frac{\partial_{k}\alpha}{2}&=&-\frac{\cos2\theta}{\sin\theta}\left(A_{12}^{R}\cos\eta +A_{12}^{I} \sin\eta\right) \nonumber\\
&&-\textrm{cotg}\frac{\alpha}{2} \textrm{cotg}\theta \left(A_{12}^{R}\sin\eta
- A_{12}^{I}\cos\eta\right)+\cos\theta (A_{11}-A_{22})
\\
\partial_{k}\theta&=&\frac{\cos\theta \sin\alpha}{\sin^2(\alpha/2)} ( A_{12}^{R} \cos\eta + A_{12}^{I}\sin\eta)\nonumber\\
&&+\frac{\cos\alpha}{\sin^2(\alpha/2)} (A_{12}^{R}\sin\eta-A_{12}^{I} \cos\eta)
-\textrm{cotg}\frac{\alpha}{2} \sin\theta (A_{11}-A_{22})
\label{eq:theta}
\end{eqnarray}
\begin{equation}
\partial_{k}\chi= 0\,,\qquad \partial_{k}\varphi=0
\label{eq:chi-phi}
\end{equation}
where we have defined $\eta\equiv\varphi-\chi$, with $\partial_{k}\eta=0$.
The solution of (\ref{eq:chi-phi}) is $\chi=\chi_{0}$, $\varphi=\varphi_{0}$.
Then, it is evident that only two equations are left, namely  (\ref{eq:alpha}) 
and (\ref{eq:theta}), with $\eta=\varphi_{0}-\chi_{0}$ playing the role of a parameter,
and with one of the angles among $\varphi_{0}$ and $\chi_{0}$ being arbitrary. We then can choose $\chi_{0}=0$ without loss of generality.
In addition, in order to conform to (\ref{eq:u-periodicity}), the angles $\alpha$ and $\theta$ must satisfy the following periodicity conditions
\begin{eqnarray}
\label{eq:alpha-period}
\alpha(-k_{B})&=&\alpha(k_{B}) + 4\ell\pi\\
\theta(-k_{B})&=&\theta(k_{B})+2\ell'\pi
\label{eq:theta-period}
\end{eqnarray}
with $\ell,\ell'$ integers which can be taken $\ell=\ell'=0$ without loss of generality. 
This set of equations,  (\ref{eq:alpha}), (\ref{eq:theta}), (\ref{eq:alpha-period}), and 
(\ref{eq:theta-period}) can be solved as discussed in \ref{sec:gaugetransf}.
The resulting transformation is an $SU(2)$ matrix $S(k)$ of the form (\ref{eq:umatrix}) with $\chi=0$ and $\varphi$ constant. 

Summarizing, the procedure to make ${\tilde \Omega}$ vanish can be divided in two steps:
\textit{(a)} a gauge transformation of type II to make $\Omega_{OD}$ vanish (without 
specific requirements on the transformation of the diagonal elements $A_{nn}(k)$);
\textit{(b)} a set of two gauge transformation of type I (one for each band) that makes $\Omega_{D}$ vanish without affecting $\Omega_{OD}$. Therefore, the full transformation can be decomposed in the following way
\begin{equation}
U_{nm}(k)= e^{\displaystyle{i\phi_n(k)}}S_{nm}(k)
\label{eq:decomposition}
\end{equation}
It is straightforward to verify that (\ref{eq:decomposition}) can be cast again in the form (\ref{eq:u0matrix}) by properly
redefining the parameters $z_1,z_3,r$.

\subsection{Examples of MLWFs}
\label{sec:examples}

Here we present some example of \textit{generalized} MLWFs obtained with the transformation discussed above, and we compare them with the \textit{single band} MLWFs. The latter can be obtained by using just the gauge transformation of type I, and correspond to the exponentially decaying Wannier functions discussed by Kohn \cite{kohn,marzari}. Both gauge transformations are solved by using the representation of Bloch functions in $k$-space, by means of the numerical methods discussed in \ref{sec:numerics} and \ref{sec:gaugetransf}. 

\begin{figure}[t]
\centerline{\includegraphics[width=0.6\columnwidth]{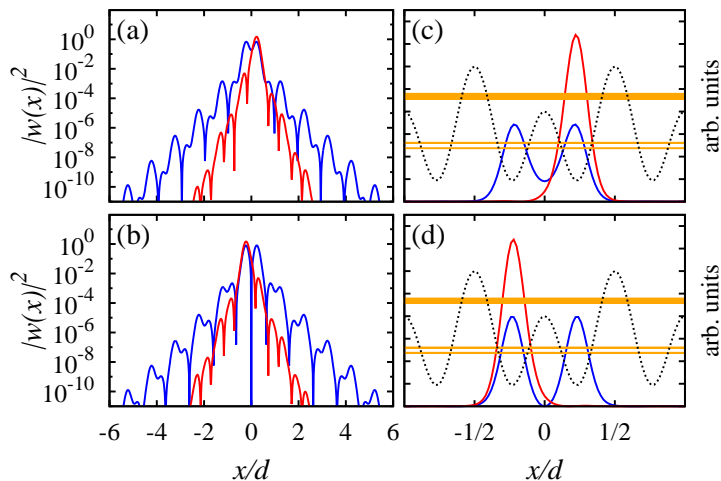}}
\caption{Plot of the density of the two lowest \textit{single band} (blue lines) and \textit{generalized} (red lines) MLWFs, in log (a,b) and linear scale (c,d). The dotted line in (c,d) represent the potential, while the horizontal orange stripes are the first three Bloch bands (on the same scale of the potential). Here $V_{1}=10E_{R}$, $V_{2}=20E_{R}$, $\theta_{0}=\pi/2$.}
\label{fig:wannier}
\end{figure}
\begin{figure}
\centerline{\includegraphics[width=0.6\columnwidth]{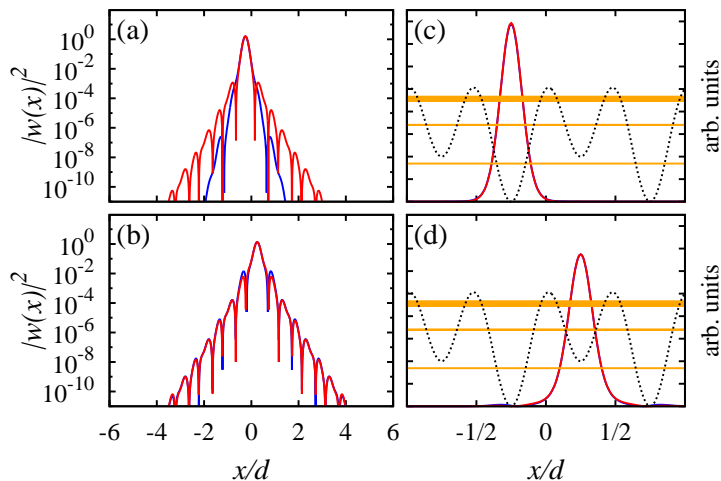}}
\caption{Same of figure \ref{fig:wannier} for $\theta_{0}=0$.}
\label{fig:wannier0-10}
\end{figure}
\begin{figure}
\centerline{\includegraphics[width=0.6\columnwidth]{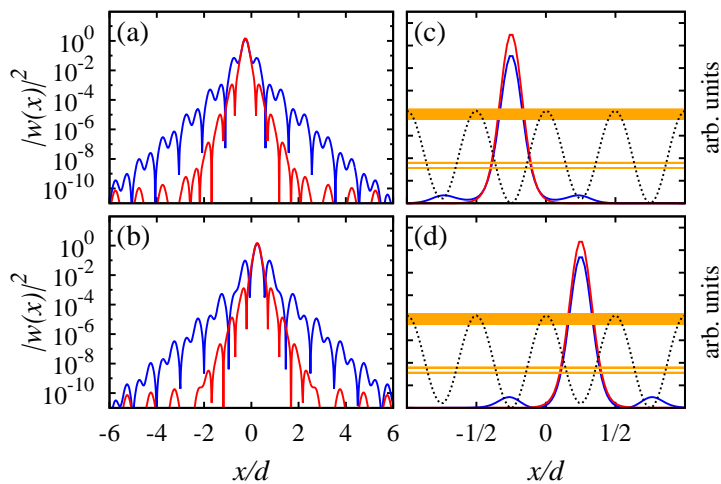}}
\caption{Same of figure \ref{fig:wannier} for $\theta_{0}=0$, $V_{1}=1E_{R}$.}
\label{fig:wannier0-1}
\end{figure}

In figure \ref{fig:wannier} we show the case of a symmetric double well, $\theta_{0}=\pi/2$.
In this case the \textit{single band} MLWFs have the same symmetry of the potential \cite{kohn}, therefore occupying both wells in the unit cell. Instead, each of the \textit{generalized} MLWFs nicely localizes in one of the sub-wells, as wanted
\footnote{We remark that in this symmetric case one could construct a set of localized functions resembling the \textit{generalized} MLWFs by considering symmetric and antisymmetric combinations of the single band MLWFs. These combinations (not shown) nicely reproduce the bulk density of the \textit{generalized} MLWFs when each unit cell can be regarded as a single double-well (that is, large barriers at the cell borders). However, this approximation does not work in the region of the tails (those of the \textit{generalized} MLWFs decaying much faster), therefore not being reliable for computing the tunnelling coefficients. It is also possible to prove analytically that this approximation implies $J_{AB-}=J_{AB+}$, that is manifestly in contradiction with the definition of the tunneling coefficients given in figure \ref{fig:doublewell}. Nevertheless, such wave functions may be useful for discussing the boundary conditions for  (\ref{eq:alpha})-(\ref{eq:chi-phi}), see \ref{sec:gaugetransf}.}.

Then, in figures \ref{fig:wannier0-10}, \ref{fig:wannier0-1} we show two cases of an asymmetric double well, here for $\theta_{0}=0$. These figures are useful for illustrating the role of the band gaps. The case in figure \ref{fig:wannier0-10} corresponds to a band gap between the first two bands that is larger than the one between the second and third band;
in this case both the \textit{single band} MLWFs are already localized within a single well,
the one in panels (b,d) having a small residual amplitude around the neighbouring wells. The \textit{generalized} MLWFs do not differ much from the former in linear scale, the effect of the gauge mixing transformation being a reduction of the lateral lobes of the \textit{single band} MLWF in (b,d), but the price to pay is a substantial increase of the width of the other one in log scale, see panel (a). Actually, in this case composite band approach is not fully justified, due to the large band gap between the first and second band, and one should 
consider the structure of the upper bands. This however goes beyond the scope of this work.

Instead, figure \ref{fig:wannier0-1} shows a case still for $\theta_{0}=0$ but where an almost degeneracy between the two minima (that corresponds to a quasi degeneracy of the lowest two bands) is restored thanks to a lower value of $V_{1}$ (here $V_{1}=1$ instead of $10$ as in the former figure). Here the advantage of using the \textit{generalized} MLWFs is clearly visible.

\section{Tight binding regimes and tunnelling coefficients}
\label{sec:regimes}

\begin{figure}[t]
\centerline{
\includegraphics[width=0.45\columnwidth]{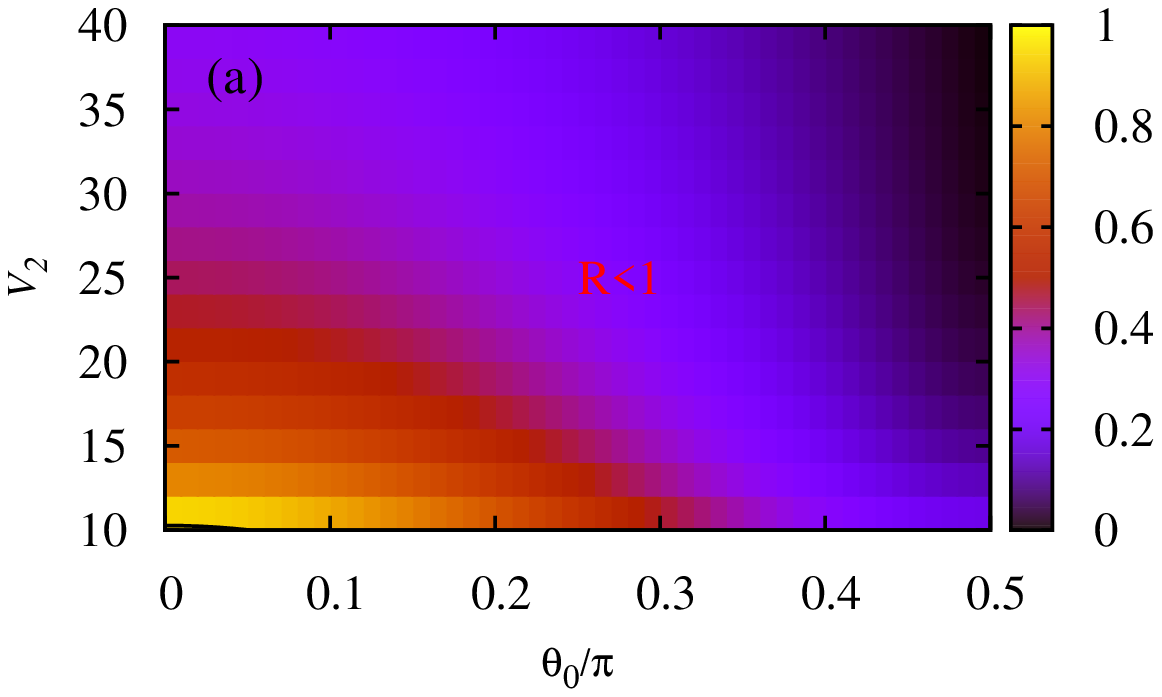}
\includegraphics[width=0.45\columnwidth]{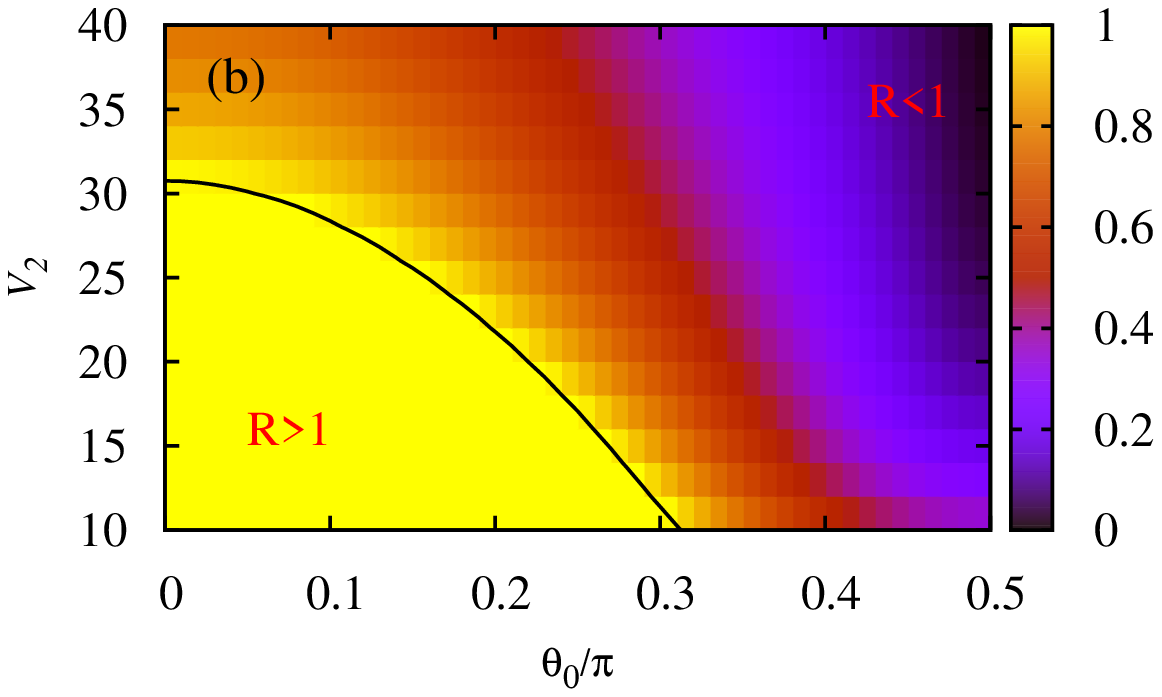}}
\caption{Density plot of the ratio between the second and first band gaps, $R\equiv\varepsilon_{g_{12}}/\varepsilon_{g_{23}}$ as a function of $\theta$ and $V_{2}$, for $V_{1}=5 E_{R}$ (a), and  $V_{1}=10E_{R}$ (b).  The black line corresponds to $R=1$. The colour scale is saturated at $R=1$.}
\label{fig:gap}
\end{figure}

In this section we will discuss the features of the \textit{composite band} approach, by comparing the predictions  of the \textit{full} and \textit{nearest-neighbour} versions of the model, and discussing the behaviour of the tunnelling coefficients and the other tight-binding parameters.  

We recall that the potential is characterized by three parameters, $V_{1}/E_{R}$, $V_{2}/E_{R}$ and $\theta_{0}$. The latter can be restricted to the range $[0,\pi/2]$ without loss of generality, see figure \ref{fig:pot}.  As for the potential amplitudes, we chose $10\le V_{2}/E_{R}\le 40$ in order to fulfil the tight binding regime, in a feasible range for current experiments with ultra cold atoms \cite{bloch}. In addition, for having both wells deep enough in order to have the corresponding MLWFs mostly localized within each well, $V_{1}$ should not exceed $V_{2}/2$.

As we have anticipated in the previous section, in principle the \textit{composite band} approach is justified in a situation of quasi degeneracy between the first two Bloch bands, that is when their separation $\varepsilon_{g_{12}}$ is much smaller that the band-gap $\varepsilon_{g_{23}}$ between the second and third band. 
Therefore, it is convenient to define the ratio $R\equiv\varepsilon_{g_{12}}/\varepsilon_{g_{23}}$, whose behaviour is shown as a density plot in figure \ref{fig:gap} as a function of $\theta_{0}$ and $V_{2}$, for different values of $V_{1}$. As a matter of fact, we find that the composite band approach 
provides a basis of functions well localized in each of the two sub-wells and correctly reproduces the lowest two Bloch energy bands even up to $R\approx1$, in the parameter range discussed above.

In order to characterize the level of fidelity in reproducing the exact single particle Bloch spectrum (that can be readily computed as discussed in \ref{sec:numerics}), we define the following quantity 
\begin{equation}
\delta \varepsilon_{n} \equiv \frac{1}{\Delta{\varepsilon}_{n}}\sqrt{\frac{d}{2\pi}\int_{\cal{B}} dk [\varepsilon_{n}(k)-\varepsilon_{n}^{tb}(k)]^{2}}
\end{equation}
that represents the ratio of the quadratic spread between the exact Bloch spectrum $\varepsilon_{n}(k)$ and that of the tight binding hamiltonians
 (\ref{eq:fullH}) and (\ref{eq:H0-cb}), to the bandwidth $\Delta{\varepsilon}_{n}\equiv (\varepsilon_{n}^{max}-\varepsilon_{n}^{min})$. We also notice that the formula (\ref{eq:ener-tb}) for $\varepsilon_{n}^{tb}(k)$ provides the same numerical result of the expression (\ref{energymlwf}) for the energy bands in terms of the gauge transformations, and that these formulas have a better accuracy in reproducing the exact Bloch spectrum than the single band expression (\ref{eq:ener-tb}), especially 
in the region close to the symmetric case $\theta_{0}=\pi/2$.

\begin{figure}
\centerline{
\includegraphics[width=0.4\columnwidth]{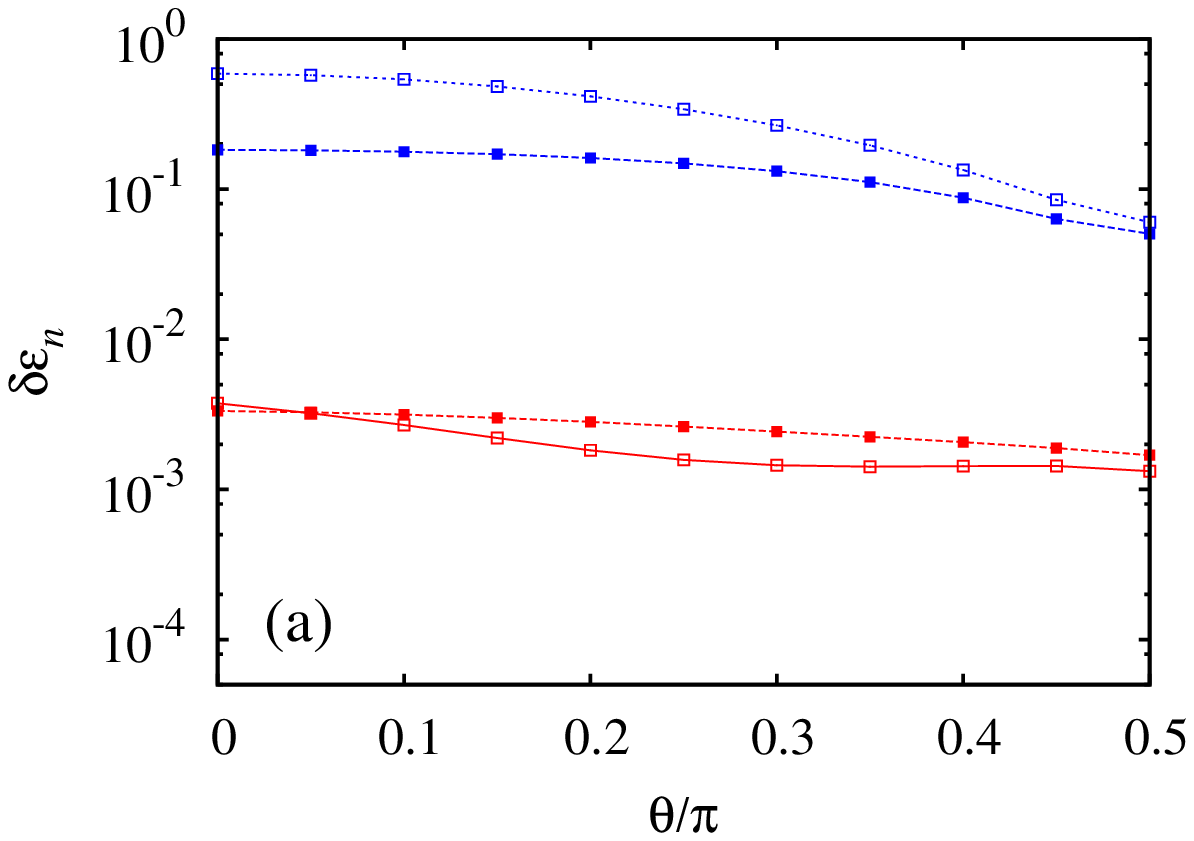}
\includegraphics[width=0.4\columnwidth]{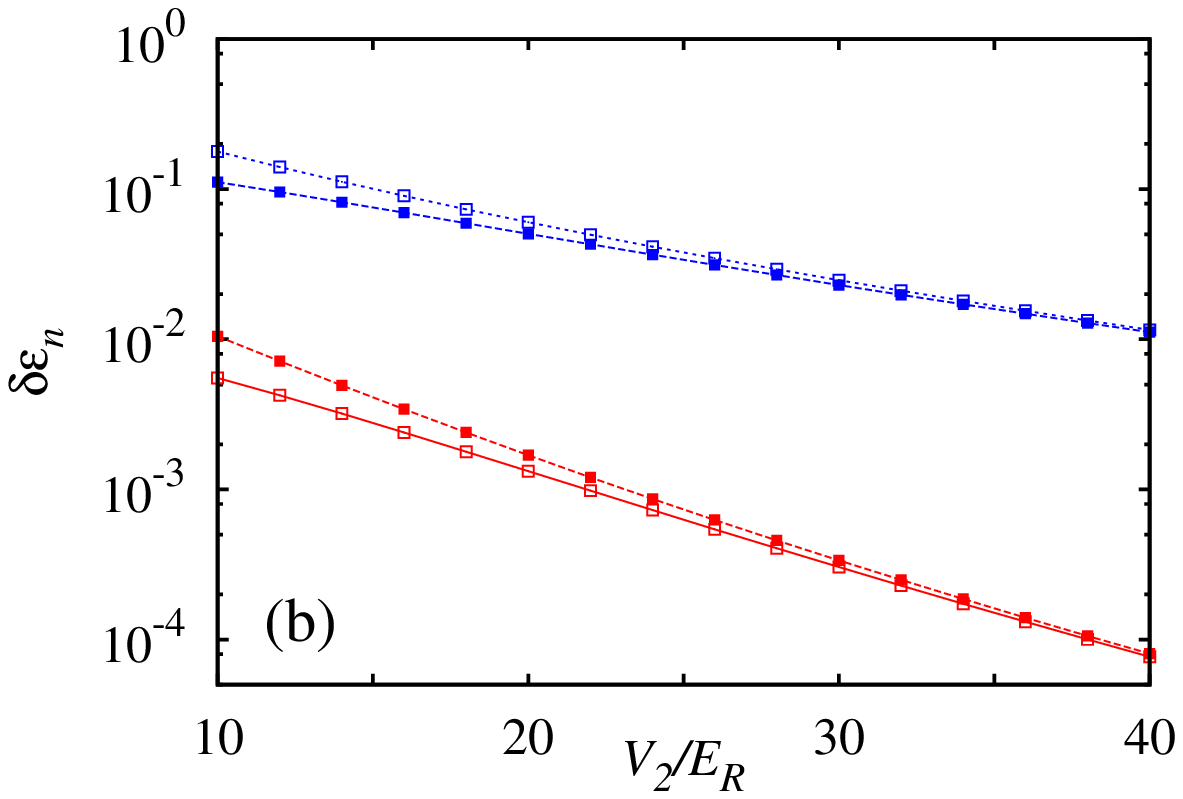}
}
\caption{Plot of the quantity $\delta \varepsilon_{n}$ (see text) for $V_{1}=5$, as a function of $\theta_{0}$ for $V_{2}=20$ (a), and as
 a function of $V_{2}$ for $\theta_{0}=\pi/2$ (b). Red symbols: \textit{full} tight-binding model; Blue symbols: \textit{nearest-neighbour}
 approximation. Empty squares: first band; Solid squares: second band. These figures refer to a horizontal and vertical cut in the left panel of figure \ref{fig:gap}.}
\label{fig:de}
\end{figure}

The quantity $\delta \varepsilon_{n}$  is shown in figure \ref{fig:de} for $V_{1}=5$, as a function of  $\theta_{0}$  ($V_{2}=20$) and $V_{2}$ ($\theta_{0}=\pi/2$), in the left and right panels respectively. These figures refer to a horizontal and a vertical cut in the left
 panel of figure \ref{fig:gap}, and confirm that in the regime $R\lesssim1$ the \textit{full} tight binding model reproduces the exact 
energies with great accuracy. As for the \textit{nearest neighbour} approximation commonly used in the literature, in general this is 
not the case, as it works reasonably only for $R\lesssim0.1$, that is for $\theta_{0}\simeq\pi/2$ and large $V_{2}$. Therefore, attention
 must be payed to the regimes where it is allowed to neglect some of the next-to-nearest tunnelling terms.

\subsection{Tunnelling coefficients}

Let us now consider the explicit expressions for the onsite energies and the tunnelling coefficients in  (\ref{eq:fullH}). They 
 can be expressed in terms of the gauge transformations discussed in the previous section as (see  (\ref{eq:wanniergen}) and 
(\ref{eq:decomposition}))
\begin{eqnarray}
E_{\nu}&=&
{\frac{d}{2\pi}}\int_{\cal B} d{k}\sum_{m=1}^{2}|S_{\nu m}(k)|^{2}\varepsilon_{m}(k)
\\
J_{\nu}&=&
-{\frac{d}{2\pi}}\int_{\cal B} d{k}e^{-ikd}\sum_{m=1}^{2}|S_{\nu m}(k)|^{2}\varepsilon_{m}(k)
\\
T_{AB}&=&
-{\frac{d}{2\pi}}\int_{\cal B} d{k} 
~ e^{i\Delta\phi(k)}\sum_{m=1}^{2}S_{1m}^{*}(k)S_{2m}(k)\varepsilon_{m}(k)
\\
J_{AB_{\pm}}&=&
-{\frac{d}{2\pi}}\int_{\cal B} d{k} 
~ e^{i(\Delta\phi(k)\mp kd)}\sum_{m=1}^{2}S_{1m}^{*}(k)S_{2m}(k)\varepsilon_{m}(k)
\end{eqnarray}
with $\nu=A,B=1,2$ and $\Delta\phi(k)=\phi_{2}(k)-\phi_{1}(k)$.
We recall that all these terms are relevant for the \textit{full} tight-binding model,
while the \textit{nearest-neighbour} approximation commonly used in the literature corresponds to retain just the contribution 
of $E_{\nu}$, $T_{AB}$ and $J_{AB_{-}}$.
We also notice that all the above parameters are \textit{gauge dependent}; $E_{\nu}$ and $J_{\nu}$ depend just on the gauge mixing 
transformation; $T_{AB}$ and $J_{AB_{\pm}}$ on the whole gauge transformation.

\begin{figure}
\centerline{
\includegraphics[width=0.4\columnwidth]{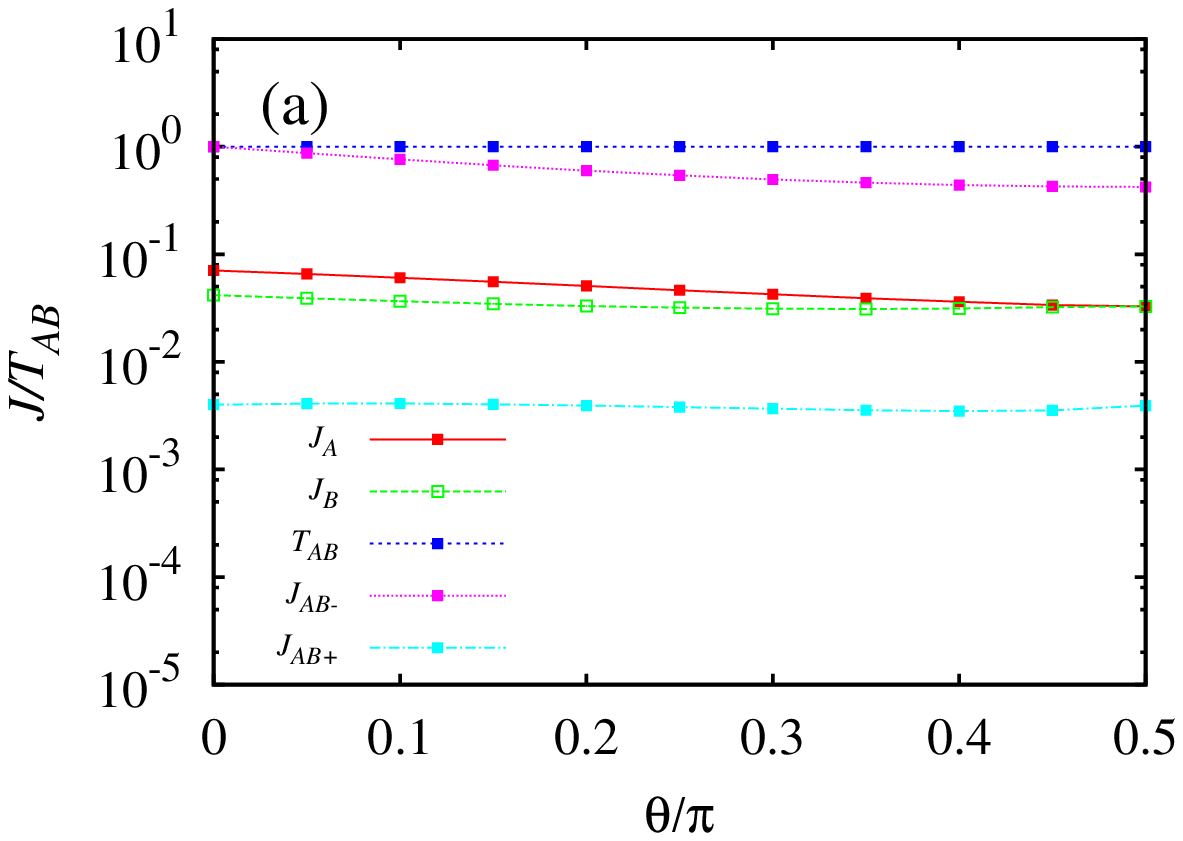}
\includegraphics[width=0.4\columnwidth]{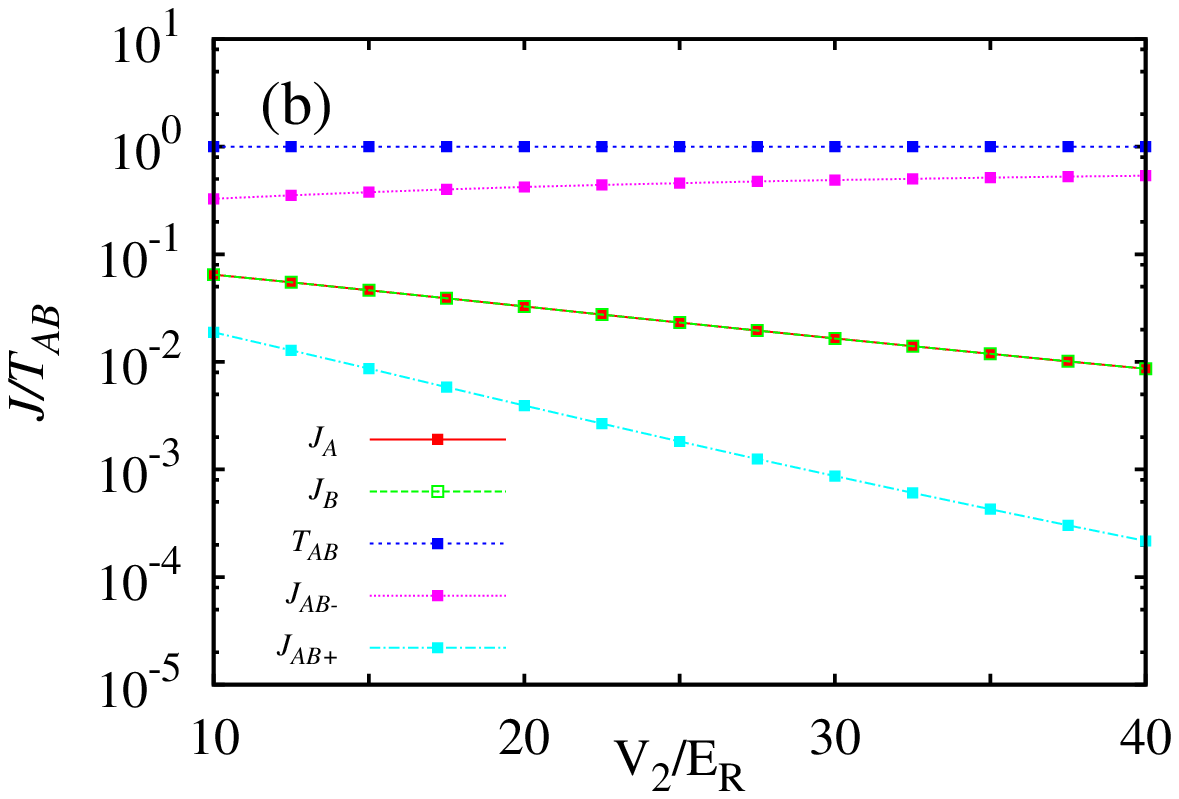}}
\caption{Plot of the tunnelling coefficients (in modulus, rescaled to $T_{AB}$) for $V_{1}=5$, as a function of $\theta_{0}$ for $V_{2}=20$ (a), 
and as a function of $V_{2}$ for $\theta_{0}=\pi/2$ (b). These figures refer to a horizontal and vertical cut in the left panel of figure \ref{fig:gap}}
\label{fig:tunnel}
\end{figure}
\begin{figure}
\centerline{
\includegraphics[width=0.4\columnwidth]{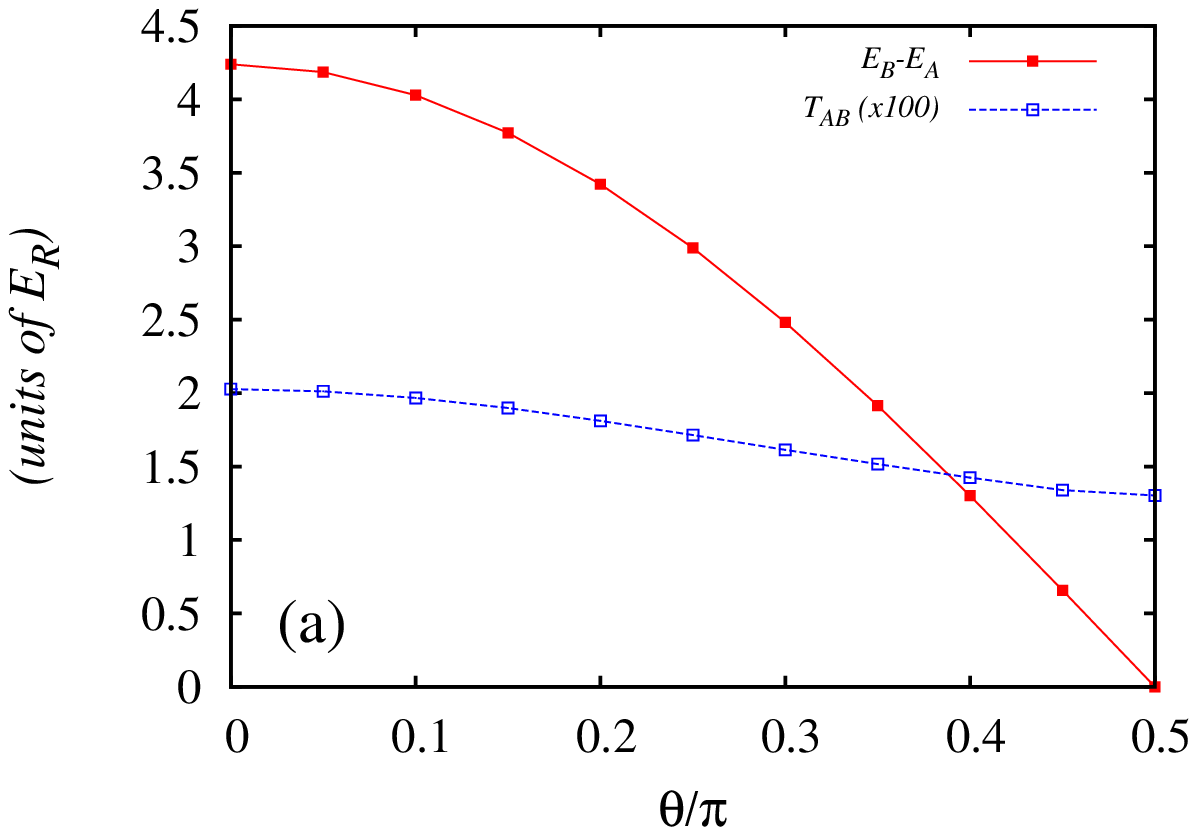}
\includegraphics[width=0.4\columnwidth]{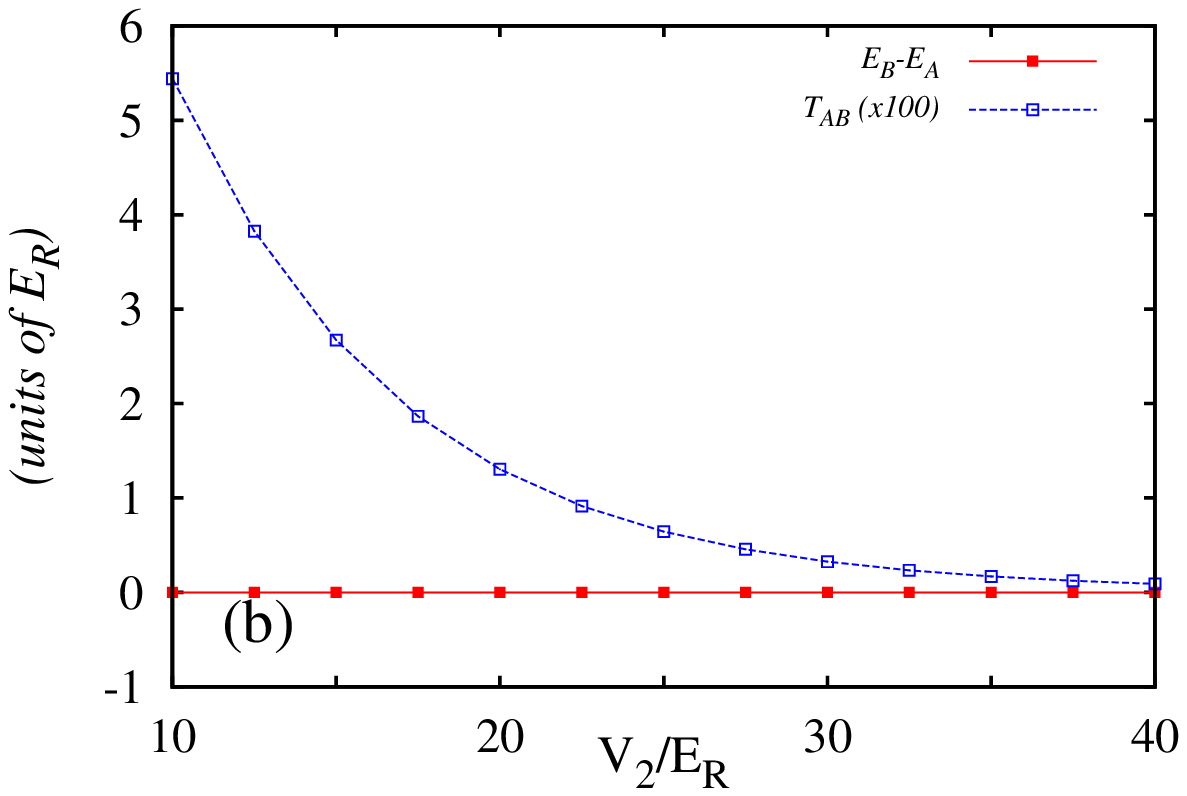}}
\caption{Plot of the modulus of $T_{AB}$ ($\times100$) and of the onsite energy difference $E_{B}-E_{A}$, for the same parameters of the previous figure.}
\label{fig:tab}
\end{figure}

It is worth mentioning that in the {\it parallel transport} gauge only the modulus of the tunnelling coefficients is univocally defined. 
However, as it is briefly 
discussed in Appendix B, there is the freedom to choose them real.

The behaviour of the tunnelling coefficients (in modulus, rescaled to $T_{AB}$) is shown in figure \ref{fig:tunnel} as a function of 
$\theta_{0}$ and $V_{2}$  (for $V_{2}=20$ and $\theta_{0}=\pi/2$, respectively; here $V_{1}=5$), whereas the absolute variation of 
$T_{AB}$ and of the onsite energy difference $E_{B}-E_{A}$ is shown in figure \ref{fig:tab}. The former figure reveals that the dependence 
on $\theta_{0}$ is rather weak, the only notable effect being that at $\theta_{0}=0$ (where all the maxima are degenerate) $T_{AB}=J_{AB+}$, whereas
at $\theta_{0}=\pi/2$ we have $J_{A}=J_{B}$ (and $E_{A}=E_{B}$, as in this case the minima are degenerate). Instead, the 
tunnelling coefficients $J_{A}$, $J_{B}$ and $J_{AB+}$ are substantially suppressed with respect to $T_{AB}$ and $J_{AB-}$ as $V_{2}$ is increased, 
and consequently the results of the nearest-neighbour approximation improve. Also,  
figure \ref{fig:de}a shows that the bands energies obtained with the nearest-neighbour
approximation better reproduce the exact results by approaching $\theta_0=\pi/2$ in spite of the very weak variation of all the tunnelling coefficients 
shown in figure \ref{fig:tunnel}. Actually, a detailed analysis reveals that these results are due to the reduction of the band gap
$\varepsilon_{g_{12}}$ and to the increase of the lowest bands widths $\Delta\varepsilon_{1,2}$ when going from $\theta_0=0$ towards $\theta_0=\pi/2$. 
Both from  (\ref{eq:ener-tb}) and  (\ref{energymlwf}) it is possible to realize that the relative weight of the terms containing $J_A$ and $J_B$ 
becomes consequently less relevant.

\begin{figure}
\centerline{
\includegraphics[width=0.4\columnwidth]{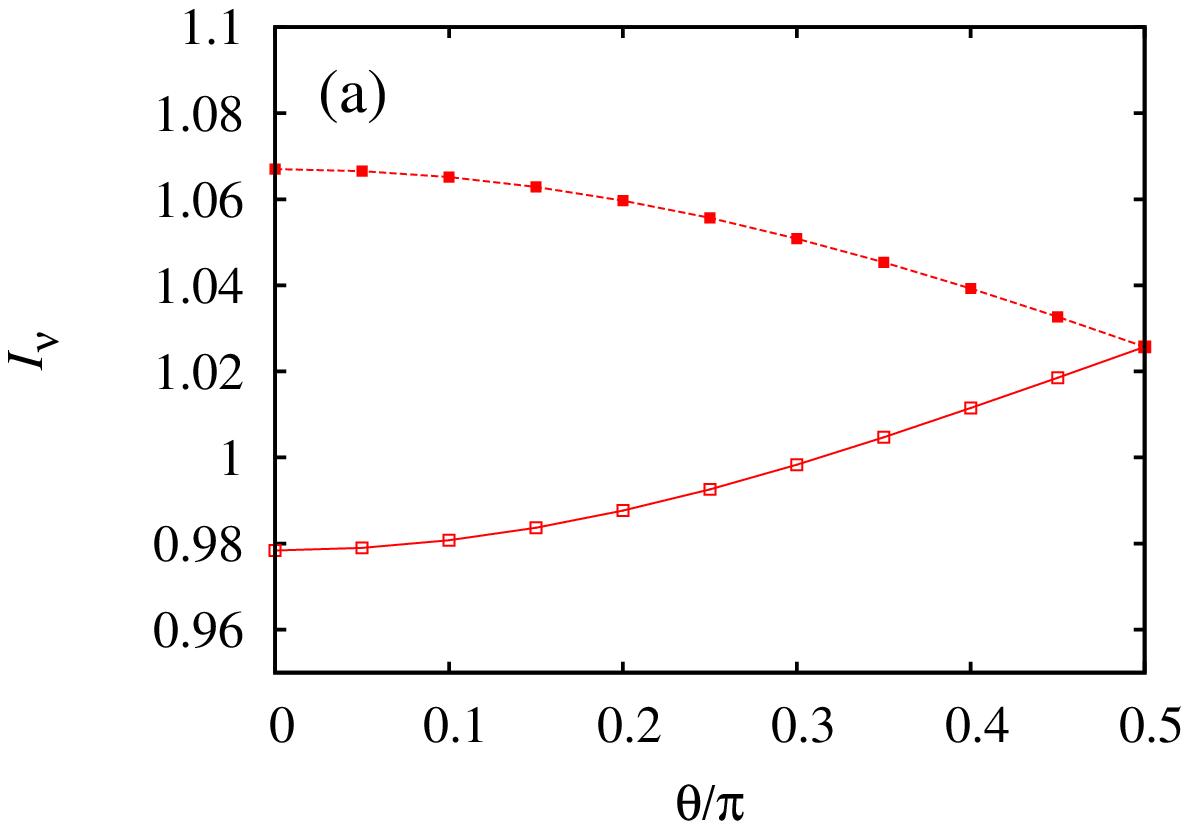}
\includegraphics[width=0.4\columnwidth]{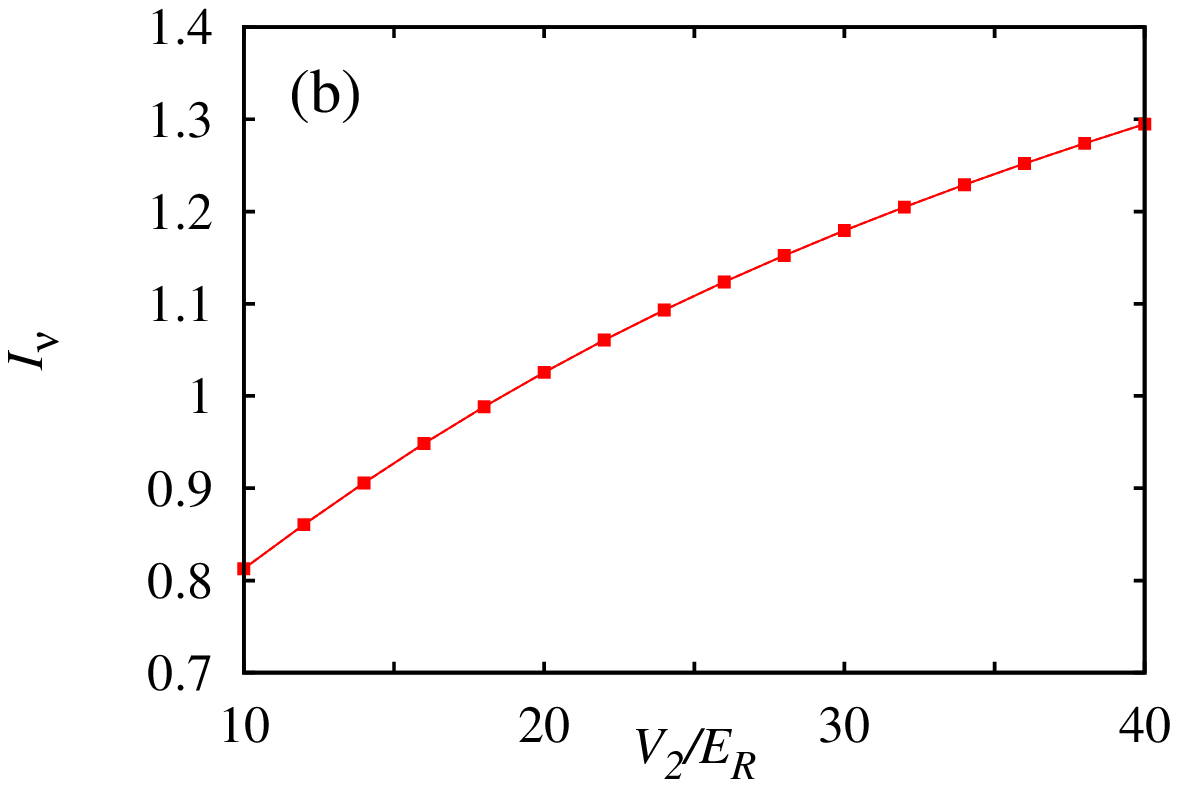}}
\caption{Plot of the rescaled interaction term $I_{\nu}$, for $V_{1}=5$, as a function of $\theta_{0}$ for $V_{2}=20$ (a), and as 
a function of $V_{2}$ for $\theta_{0}=\pi/2$ (b). Empty squares: first band; Solid squares: second band. In the right panel the two lines are superimposed owing to the double well symmetry.}
\label{fig:w4}
\end{figure}

Finally, in figure \ref{fig:w4} we consider the behaviour of the rescaled interaction term (see  (\ref{eq:Hint-cb})),
$I_{\nu}\equiv U_{\nu}/g=\int\!d x \left|\tilde{w}_{j_{\nu}} (x)\right|^4$, again as a function of $\theta_{0}$ and $V_{2}$, for the parameters 
of the preceding figures. This term is also called inverse participation ratio (IPR), and represent the number of occupied lattice sites in the case of a model defined on a discrete lattice (see e.g. \cite{modugno,ingold}). In the present continuous case, values of $I_{\nu}$ close or greater 
than $1$ imply that the MLWFs are localized on a distance smaller than the lattice spacing\footnote{In the regime of parameter considered here, the next-to-leading interaction terms - see (\ref{eq:Hint-full}) - are suppressed by about 2 orders of magnitude.}.

\section{Conclusions}
\label{sec:conclusions}

We have presented a method for constructing a set of maximally localized Wannier functions (MLWFs)
for a one-dimensional periodic potential with a double-well per unit cell. Starting from the minimal spread requirement by Marzari \textit{et al.} \cite{marzari}, we have designed a two-step gauge transformation specific for a composite two band system, consisting a set of ordinary differential equations with periodic boundary conditions. In the tight binding regime, this method provides a proper set of MLWFs when both wells are sufficiently deep, provided that the band gap between the first two bands is larger than the one between the second and third band. 
By using these MLWFs one can map the continuous hamiltonian onto a tight binding model, whose 
coefficients are expressed in terms of the above gauge transformations, providing a direct way to relate the tight binding coefficients to the experimental parameters characterizing the continuous potential. 
We have also shown that in general, in the range of typical experimental parameters, it is necessary to retain all possible tunnelling couplings between the different sub-wells of neighbouring cells (\textit{full} model), whereas the usual \textit{nearest neighbouring} approximation used in the literature is justified only for $V_2/E_{R},V_2/V_1\gg 1$ and $\theta_0\simeq\pi/2$. The method can be applied also in higher dimensions, in case of separable lattice potentials.

\ack

We acknowledge useful discussion with D Trypogeorgos, R Franzosi, and I L Egusquiza. This work has been supported by the UPV/EHU under program UFI 11/55.

\begin{appendix} 
\section{Numerics: discretization in momentum space}
\label{sec:numerics}

In order to address the problem from the numerical point of view, we use the following 
truncated Fourier expansion of the Bloch eigenfunctions
\begin{equation}
\label{eq:f-expansion}
u_{nk}(x)=\frac{1}{\sqrt{2\pi}}\sum_{\ell=-N}^{N}c_{n\ell}(k)e^{i2k_{B}\ell x}
\end{equation}
with $\sum_{\ell}|c_{n\ell}|^2=1$. The eigenvalue equation $H|u_{nk}\rangle= \varepsilon_{n}(k)|u_{nk}\rangle$ is then mapped onto a matrix eigenvalue equation for the coefficients $c_{n\ell}(k)$, and can be solved by means of standard linear algebra routines, like LAPACK.
Then, the Berry connections can be written as 
\begin{equation}
A_{nm}(k)=i\sum_{\ell}c^{*}_{n\ell}(k)\frac{\partial c_{m\ell}(k)}{\partial k}
\end{equation}
and can be computed on a $k$-grid with spacing $\Delta k$
\footnote{Typically have used from $500$ to $10000$ grid points.}, by means of \textit{symmetric two-point derivative}
\begin{equation}
\frac{\partial c_{m\ell}(k)}{\partial k}\simeq \frac{c_{m\ell}(k+\Delta k)-c_{m\ell}(k-\Delta k)}{2\Delta k}
\label{eq:c-deriv}
\end{equation}
 by using the periodicity condition $c_{n\ell}(k+2k_{B})=c_{n\ell +1}(k)$, that follows from the fact that the $\psi_{nk}(x)$ are periodic in quasimomentum with period $2k_{B}$, or equivalently that 
\begin{equation}
u_{nk+2k_{B}}(x)=e^{-i2k_{B}x}u_{nk}(x)
\label{eq:uperiodicity}
\end{equation}

Since the diagonalization at each point in $k$-space is uncorrelated from that at neighbouring $k$s, in general the diagonalization routines return a set of coefficients $\bar{c}_{m\ell}(k)$ with an arbitrary phase factor $e^{i\varphi_{nk}}$ that cannot be controlled a priori, and that affects their differentiability in 
(\ref{eq:c-deriv}). However, a set of smooth (differentiable) $c_{m\ell}(k)$ can be obtained as follows. We notice that in $x=0$ the periodicity condition (\ref{eq:uperiodicity}) reads $u_{nk+2k_{B}}(0)=u_k(0)$. Therefore, given a set of Bloch eigenfunctions $\bar{u}_{nk}(x)$ generated numerically (affected by the same spurious phase factors $e^{i\varphi_{nk}}$ of the $\bar{c}_{n\ell}(k)$, see (\ref{eq:f-expansion})) it is possible to define a new set of smooth Bloch eigenfunctions $u_{nk}(x)$ by fixing all the phases to zero at $x=0$ (i.e. by imposing them to be real in $x=0$), 
\begin{equation}
u_{nk}(x)=\bar{u}_{nk}(x)\exp(-i\arg(\bar{u}_{nk}(0)))
\end{equation}
therefore fixing the spurious phases.
Formally, this corresponds to a diagonal gauge transformation generated by $\phi_{n}(k)=-\arg(\bar{u}_{nk}(0))=-\arg\left(\sum_{\ell}\bar{c}_{n\ell}(k)\right)$. Since $\phi_{n}(k)$ does not depend on $\ell$, the $c_{n\ell}(k)$ transform in the same way
\begin{equation}
c_{n\ell}(k)=\bar{c}_{n\ell}(k)\exp\left(-i\arg\left(\sum_{\ell}\bar{c}_{n\ell}(k)\right)\right)
\end{equation}
and this can be used as a prescription for ensuring the smoothness of the $c_{n\ell}(k)$ and $u_{nk}$ as a function of $k$.

\section{Numerics: gauge transformations}
\label{sec:gaugetransf}

\textit{Diagonal transformation}.
The diagonal gauge transformation of type I is obtained by solving the ordinary differential equation for $\phi_{n}(k)$ in  (\ref{eq:gauge-a}) by using a 4th-order Runge-Kutta (RK4) integrator \cite{press}, with initial condition $\phi_{n}(-k_{B})=0$. A single run returns a (single band) diagonal spread $\Omega_{Dn}=\epsilon>0$, with $\epsilon\ll1$; the procedure can be iterated in order to achieve the desired accuracy $\epsilon_{min}$, $\Omega_{Dn}<\epsilon_{min}$ (typical value: $\epsilon_{min}=10^{-18}\div10^{-25}$).

\textit{Gauge mixing transformation: method 1.}
The solution of  (\ref{eq:alpha}) and (\ref{eq:theta}) with periodic boundary conditions (\ref{eq:alpha-period}), (\ref{eq:theta-period}) (here we fix $\ell=\ell'=0$) can be achieved by using two nested \textit{shooting algorithms}, adapting the discussion in \cite{press} to the case considered here. As a first step (``inner loop''), we fix a trial value for $\eta$, and we chose an arbitrary initial value $\alpha(-k_{B})=\alpha_{0}$ (avoiding the singular value $\alpha_{0}\neq\pi$). Then we impose the periodic boundary conditions $\alpha(-k_{B})=\alpha(k_{B})$ with a shooting method that uses the initial value $\theta(-k_{B})$ as free parameter, that is varied in order to match the required boundary conditions on $\alpha$ \cite{press}. The differential equations are integrated with the same RK4 integrator mentioned above. The second step (``outer loop'') consists in a similar shooting algorithm, that uses $\eta$ as a free parameter trying to match the periodic boundary conditions on $\theta$, $\theta(-k_{B})=\theta(k_{B})$, without affecting those on $\alpha$ (that are constrained by the ``inner loop''). With this method it is possible to reduce $\Omega_{OD}$ by several orders of magnitudes, down to $\simeq 10^{-12}$, (the actual value depends on the number of grid points and on the region of parameters). We also find that the method is more efficient if it is preceded by the diagonal transformation discussed above.

\textit{Gauge mixing transformation: method 2.}
In order to solve  (\ref{eq:alpha}), (\ref{eq:theta}), we have also employed another method that originates from the observation that
in the case of two degenerate symmetric minima in the elementary cell ($\theta_0=\pi/2$),
simple symmetric and antisymmetric combinations of Bloch eigenstates give rise to two Wannier functions located each at one of the two degenerate minima (though they are not the maximally localized ones). 
In this situation, we have verified that initial conditions for the matrix $S_{nm}(\pm k_{B})$
- corresponding to symmetric and antisymmetric combinations of Bloch eigenstates -
allow to obtain the periodic solutions minimizing $\Omega_{OD}$ which coincide both integrating from $-k_{B}$ to  $k_B$ and vice-versa. 
By smoothly changing $\theta_{0}$, we start from the same initial conditions and then we iterate
the integration of   (\ref{eq:alpha}), (\ref{eq:theta}) by taking as initial angles, at each step, the mean values of the  final values  of the preceding integration (at the two boundaries of the BZ).
 The procedure rapidly converges  with $\Omega_{OD}$ reduced by several order of magnitude (this depends on the number of grid points and on the region of parameters, as for the previous method). 
In particular, for $\theta_0=\pi/2$ (symmetric double well) we have always found solutions with $\varphi_0=\pi/2$ and the following boundary conditions for $\alpha$ and $\theta$ 
\begin{equation}
\alpha(\pm k_B)=\theta(\pm k_B)=\pi/2.
\end{equation}
By smoothly changing $\theta_0$, the corresponding boundary conditions are smoothly connected with this set (in the range of $V_1$ and $V_2$ that we have explored). Notably, this choice yields real tunnelling coefficients (in principle this is not guaranteed, see \ref{sec:ambiguities}).

\section{Uniqueness of the physical results}

\subsection{Energy bands in the truncated expansion}

Let us start by considering the following exact expression for the Bloch spectrum 
\begin{equation}
\varepsilon_{n}(k)=\sum_{\ell}e^{ik\ell d} T_{n}(\ell)
\label{eq:singleBloch}
\end{equation}
with the coefficients $T_{n}(\ell)$ being the  expectation values of $\hat{H}_{0}$ on single band Wannier functions \cite{he}
\begin{equation}
T_{n}(\ell)\equiv \bra{w_{nj}}\hat{H}_{0}\ket{w_{nj+\ell}}={\frac{d}{2\pi}}\int_{\cal B}\!\!d{k} ~\varepsilon_{n}(k) e^{-ikd\ell}.
\label{eq:exp-w}
\end{equation}
This expression is manifestly gauge invariant. Instead, the analogous expression in terms of the generalized MLWFs depends on the gauge. In fact, with some manipulations we can write (here we are implicitly considering just the first two bands, $n=1,2$)
\begin{equation}
\varepsilon_{n}(k)=\sum_{\ell}e^{ik\ell d} {\tilde P}_{n}(k,\ell)
\label{eq:compositeBloch}
\end{equation}
with 
\begin{equation}
{\tilde P}_{n}(k,\ell)=\sum_{n'n''}U_{n''n}(k)U^{*}_{n'n}(k){\tilde T}_{n''n'}(\ell)
\end{equation}
and
\begin{equation}
{\tilde T}_{n''n'}(\ell)=\langle{\tilde w}_{n''j}|\hat{H}_{0}|{\tilde w}_{n'j+\ell}\rangle.
\end{equation}
By truncating (\ref{eq:compositeBloch}) to nearest neighbouring cells ($\ell=0,\pm1$)
and using the definitions of the onsite energies and of the tunnelling coefficients in  (\ref{eq:jnu})-(\ref{eq:jab}), we have
\begin{eqnarray}
\varepsilon_{n}(k)&=&|U_{1n}|^2 E_{A}+|U_{2n}|^2 E_{B}-2|U_{1n}|^2{\rm Re}\left(e^{ika}
 J_{A}\right)-2|U_{2n}|^2{\rm Re}\left(e^{ika}
J_{B}\right)
\nonumber\\
&&-2{\rm Re}\left[U_{1n}U^{*}_{2n}\left(T_{AB}+e^{ika}J_{AB_{+}}+e^{-ika}J_{AB_{-}}\right)\right]
\label{energymlwf}
\end{eqnarray}
with $U_{nm}(k)=e^{i\phi_{n}(k)}S_{nm}(k)$ and $S\in SU(2)$ (see Sect. \ref{sec:mlwfs}).
Though each energy $\varepsilon_{n}(k)$ in  (\ref{energymlwf}) depends on the gauge (through the the matrix $U_{nm}(k)$), it is not difficult to demonstrate that their sum, $\sum_{n=1,2}\varepsilon_{n}(k)$, is gauge invariant like the analogous expression obtained form (\ref{eq:singleBloch}). In addition, once we have chosen the {\it parallel transport gauge}, corresponding to the generalized MLWFs with minimal spread (${\tilde{\Omega}}=0$), the expression (\ref{energymlwf}) is univocally defined despite the presence of residual phase arbitrariness (see next paragraph). 

\subsection{Residual phase ambiguities}
\label{sec:ambiguities}

Let us now show that the relevant quantities are univocally defined in the {\it parallel transport gauge}, despite possible ambiguities in the procedure for obtaining the matrix $U_{nm}(k)$.
First of all, we remind that the whole procedure starts from a particular gauge, as discussed in \ref{sec:numerics}. Choosing a different initial gauge amounts to 
perform a diagonal transformation on Bloch eigenstates 
\begin{equation}
|\psi_{mk}\rangle\rightarrow e^{i\phi_{m}(k)} |\psi_{mk}\rangle
\end{equation}
being $\phi_{m}(k)$ an arbitrary  periodic real function (see Sect. \ref{sec:mlwfs}).
Consequently, the full gauge transformation matrix leading to the {\it parallel transport gauge} changes as
\begin{equation}
U_{nm}(k)\rightarrow U_{nm}(k)e^{-i\phi_{m}(k)}.
\end{equation}
Then, it is easy to see that the onsite energies and tunnelling coefficients in  (\ref{eq:jnu})-(\ref{eq:jab}) are invariant under this transformation,
as well as the energies in  (\ref{energymlwf}), 
whereas the real and imaginary parts of the MLWFs change, conserving their modulus.
The procedure to make ${\tilde\Omega}$ vanishing does not imply any other 
 $k$-dependent ambiguity on the matrix $U_{nm}(k)$.  
In addition, in the {\it{parallel transport}} gauge we still have the freedom to  perform two independent $U(1)$ gauge transformations for each band  with angles $\lambda_n$ which necessarily obey  $\partial\lambda_n/\partial k=0$. These terms appear as multiplicative factors on the rows of the matrix $U_{nm}$.
Finally, it turns out that - given a set of parameters $V_1,V_2$ and $\theta_0$ - there is not a unique set of boundary conditions for the angles $\varphi_0$, $\alpha$ and $\theta$ satisfying  (\ref{eq:alpha}), (\ref{eq:theta}) (and thus not an unique transformation $S_{nm}(\pm k_B)$ leading to $\Omega_{OD}=0$). For $\theta_0=\pi/2$ these boundary values can be fixed as discussed in \ref{sec:gaugetransf}. Different choices may also be possible.
However, it is not difficult to prove that none of these ambiguities affect either the energies in (\ref{energymlwf}), or the modulus of the MLWFs and of the tunnelling coefficients. Instead, their real and imaginary parts are not univocally defined and can be chosen somewhat arbitrarily.  In particular, real tunnelling coefficients can be obtained by using the set of boundary conditions given at the end of \ref{sec:gaugetransf}.

\end{appendix}

\section*{References}

\end{document}